\documentclass[onecolumn,groupedaddress,aps,prb,preprint,groupedaddress,amsmath,amssymb,showpacs]{revtex4}
\usepackage{graphicx}
\usepackage{dcolumn}
\usepackage{latexsym,bm}
\usepackage{longtable}

\begin{document}
\preprint{}
\title{{Error analysis of the density-matrix renormalization group algorithm for a chain of harmonic oscillators}}

\author{ Ma Yongjun£¨ÂéÓÀ¿¡£©, Wang Jiaxiang£¨Íõ¼ÓÏ飩\footnote{Corresponding author: jxwang@phy.ecnu.edu.cn,13916870533(cell phone number), 021-62233773(office phone number)},Xu Xinye£¨ÐìÐÅÒµ£©, Wei Qi£¨ÎºÆô£©}

  \affiliation{State Key Laboratory of Precision
Spectroscopy, East China Normal University, Shanghai 200062, China }

\author{Sabre Kais}

  \affiliation{Departments of Chemistry and Physics, Purdue University, West Lafayette, Indiana 47907, USA}

  \affiliation{Qatar Environment and Energy Research Institute, Qatar Foundation, Doha, Qatar}


\begin{abstract}
  We investigate the application of the density-matrix renormalization
group (DMRG) algorithm to a one-dimensional harmonic oscillator chain and
compare the results with exact solutions, aiming to improve the
algorithm efficiency. It has been demonstrated that the algorithm
can give quite accurate results if the procedure is proper
organized, for example, by using the optimized bases. The errors of calculated ground state energy and the energy gap between the ground
state and the first excited state are  analyzed, which are found to
be critically dependent upon the size of the system or the energy level structure of the studied system and the number of states targeted during
the DMRG procedure.


\end{abstract}

\maketitle

The density matrix renormalization group method (DMRG
)\cite{dmrg01,dmrg1} is well known for its high-accuracy in studying
low-dimension physical system. But in dealing with the bosonic
system, we meet great challenges due to the infinite
dimensions of the local Hilbert space. So it is unavoidable to truncate the base space in
the DMRG procedure. The question is how to minimalize the truncation errors and to find out the factors which could heavily influence the accuracy.
To acquire the answer is the basic motivation for us to carry out the
work in the letter.
  \par
   We will choose the one-dimension oscillator chain as the
model, since it has analytical solution and the errors can be
conveniently analyzed.

\begin{figure}
\includegraphics[angle=-90,width=10cm]{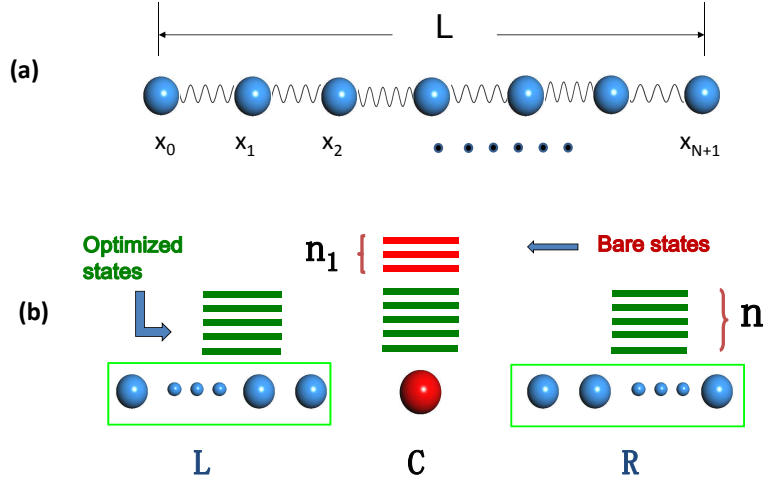}
\hfill \caption{(a) Schematic presentation of the oscillator chain
model. A chain of particles are connected by the springs  with fixed
boundary: $x_{0}=0,\quad x_{N+1}=(N+1)a$. The average distance
between neighboring particles is $a=L/(N+1)$. (b) Demonstration of the algorithm to rebuild the optimized basis. The
system block (L) and the environment block (R) are both n-dimensional,
while the central point has a dimension of $(n+n_{1})$. A group of $n_{1}$ bare base states will be fed into the whole system in each sweeping loop systematically to form a new superbolock Hamiltonian. Then the new reduced density matrix of the central site will be used to obtain the $n$ optimized states.}
\label{oscillatorchain}
\end{figure}
 \par
As shown in Fig.\ref{oscillatorchain}(a), the oscillator chain is
composed of N+2 particles connected by springs, with the fixed
boundary conditions: $x_{0}=0,\quad x_{N+1}=(N+1)a$, where a is the
average distance between neighboring particles. The Hamiltonian can
be expressed as,
\begin{equation}
\hat{H}=\sum_{i=1}^{N}[- \frac{\hbar^2}{2m}\frac{\partial^2
}{\partial x_{i}^{2}}+ \frac{k}{2}({x_{i+1}}-{ x_{i}}-a)^2,
] \label{E_interna0}
\end{equation}
in which  $\hbar$ is the Planck constant, $k$ the elastic
constant, $m$ the particle mass and $x_i$ the coordinate for the \emph{i}th particle. For convenience, the length, the mass and the energy will be scaled by $a$,
$m$ and $ka^2$ respectively, which leads to the following dimensionless Hamiltonian,
\begin{equation}
\hat{H'}=\sum_{i=1}^{N}[- \frac{
{\tilde{\hbar}}^2}{2}\frac{\partial^2 }{\partial U_{i}^{2}}+
\frac{1}{2}({ U_{i+1}}-{ U_{i}})^2 ], \label{reaction2}
\end{equation}
in which $\hat H' = \hat H/k{a^2}$, $U_{i}= x_{i}/a-i$ and $\tilde{\hbar}=\hbar\frac{1}{\sqrt{mk}}$.
The analytical form of the
dispersion relationship of the above Hamiltonian can be readily
obtained,
\begin{equation}
  \omega_{q_n}=2\cdot|\sin\frac{q_n}{2}|, \quad
q_n=\frac{n\pi}{N+1},(n= 1,2\cdots,N),
\end{equation}
from which we can obtain the ground state energy and the energy gap between the ground
state and the first excited state,
\begin{equation}
  E_{ground}=\frac{1}{2}\sum_{q_n} \tilde{\hbar} \omega_{q_n }
  = \frac{\sqrt{2}}{2}\cdot\tilde{\hbar} \cdot\frac{\sin \frac{N\pi}{4(N+1)}}
   {\sin \frac{\pi}{4(N+1)}}, \label{Eground}
\end{equation}
\begin{equation}
  \bigtriangleup E_{12}= \tilde{\hbar}\cdot\\sin \frac{\pi}{2(N+1)}.
    \label{deltE}
\end{equation}
\par
  Next we will try to obtain the above solution with DMRG method numerically. For this purpose, we need to rewrite
Eq.\ (\ref{reaction2}) into a second-quantized form with the following transformation,
\begin{eqnarray}
{U_{i}} = \frac{1}{\sqrt{2}} \frac{\sqrt{ {\tilde{\hbar}}}}
{\sqrt[4]{2}}   (\hat{a_{i}}^{\dag}+\hat{a_{i}}), \qquad
{P_{i}}= -i\tilde\hbar\frac{\partial }{\partial{U_{i}}} = -\frac{1}{\sqrt{2}}
\frac {\sqrt[4]{2}}{\sqrt{ {\tilde{\hbar}}}}
(\hat{a_{i}}^{\dag}-\hat{a_{i}}),
\end{eqnarray}
 in which $\hat{a_{i}}^{\dag}$ and $\hat{a_{i}}$ are the annihilation
 and the creation operators satisfying,  $[{\hat a_{i}},{\hat a_{j}^{\dag}}]=
 \delta_{ij}$,  $[{\hat a_{i}},{\hat a_{j}}]= 0$,
 $[{\hat a_{i}}^{\dag},{\hat a_{j}^{\dag}}]= 0$.  Then we can get the new
 Hamiltonian,
\begin{eqnarray}
\hat{H'}& = & \sqrt{2} {\tilde{\hbar}} \sum_{i=1}^{N}
\{(\hat{a_{i}}^{\dag}\hat{a_{i}}
 +\frac{1}{2} )  \}
  -\frac{\sqrt{2} {\tilde{\hbar}}}{4}\sum_{i=1}^{N}({\hat a_{i}}^{\dag}
  +{\hat a_i} )( {\hat a_{i+1}}^{\dag} +\hat a_{i+1}). \label{action4}
\end{eqnarray}


\par
The scheme of standard DMRG  can be found in many
reference\cite{dmrg01,dmrg1,dmrg2}. The main idea is to make an
effective Hamiltonian, which includes one renormalized left block (L), one renormalized right block (R), and one or two free site(s) between the blocks. The new
Hamiltonian has the same low-lying energy levels as the old ones. For a bosonic system, Fig.\ref{oscillatorchain}(b) present a schematic explanation of the algorithm.
Firstly, the bases in the local Hilbert space of the \emph{i}th site $\{ \left|n_i\right\rangle,
(n_i=1,2,3\cdot\cdot\cdot\infty) \}$ is truncated to $\{ \left|n_i\right\rangle,
(n_i=1,2,3,\cdot\cdot\cdot,m) \}$, where $\left |n_i\right\rangle$ is the particle number state called bare states. Then by standard DMRG algorithm, we keep $n$ renormalized
bases for each site and each block. It should be mentioned that unlike the fermion systems, the site bases here also need to be renormalized and truncated. Thus besides the traditional truncation of the block bases, we also require two more truncations. One is to truncate the bare bases to $m$,  which will form our working space in all the following calculations.  Normally $m$ will be chosen to be large enough to guarantee the convergence of the results. The other is to truncate the renormalized local bases up to $n$ as shown in Fig. \ref{oscillatorchain}(b). To control the above two truncations, especially the second one, plays a key role in improving the numerical accuracy. In order to minimize the
errors from the second truncation, those states above the $n$ renormalized states, which is neglected in traditional DMRG algorithm, will also be used in each sweeping loop systematically to form a new set of bases called optimized bases. For brevity, the realization details will not be repeated here and can be found in some earlier
papers \cite{F-K4,dmrg0, method1,method2,method3,method4}.

  \par

   \begin{figure}
 \includegraphics[angle=-90,width=13cm]{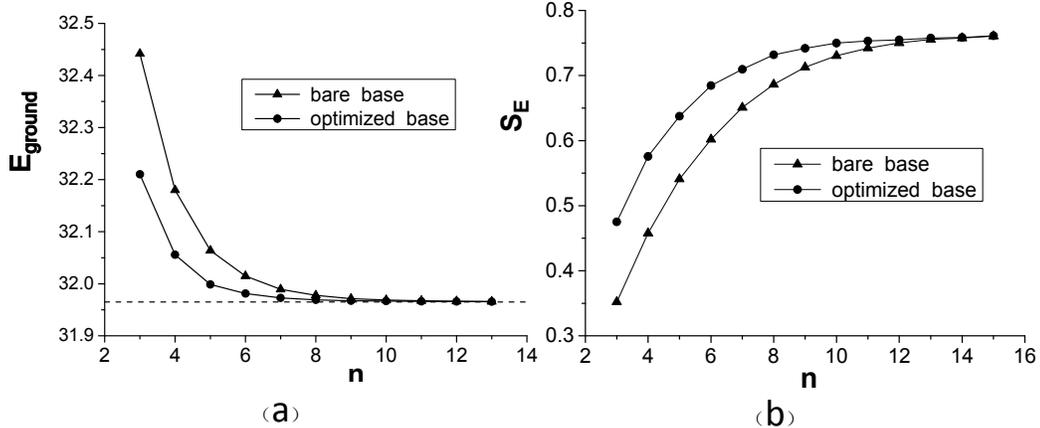}
 \caption{Dependence of ground state energy [(a)] and entanglement [(b)] calculation upon the number of renormalized bases in DMRG algorithm. The system size is $N=50$. In (a), the dotted line denotes the exact results obtained from Eq.(\ref{Eground}).}
 \label{es}
 \end{figure}

  Fig. \ref{es}(a) presents the influence of the number of the basis for each site or block upon the
ground energy for the chain size $N=50$. By compared with the exact results, it can be easily found that the more bases are used, the  more accurate results we can acquire. Moreover, the numerical results are improved
distinctly as the optimized bases are utilized.  For example, with 8 optimized-basis can give the results almost as accurate as with 10 bare bases. This is a great improvement of the calculation efficiency since it means the solution of a matrix with dimension $8\time 8\times 8=512$ instead of $10\times 10\times 10=1000$. The reason is quite clear since the optimized bases contains more contribution from higher energy levels.
  \par The above convergence with the number of the basis and the improvement with optimized bases work not only for the ground state energy, but also for the entanglement. As we know, the quantum entanglement is now considered as a potential
resource which is widely applied in the quantum communications and
computations \cite{Kais}. In the research of quantum phase transitions, it can be also taken as an order parameter due to its critical property \cite{wang2003,wang2004}.
Hence the entanglement calculation is needed in many models. Here we will use the von Neumann entropy as a measure of the entanglement,
  $S  =  -tr(\rho\log \rho )$, where $\rho$ is the density matrix and $tr$ denotes the trace, to calculate the following average
  local entanglement,\cite{entangle1,entangle2,entangle3}
\begin{eqnarray}\label{7}
  S_{E}=\frac{1}{n} \sum\limits_{i=1}^n S_i,
\end{eqnarray}
where $S_i$ is the entanglement of $i$th particle with the rest part of the chain, $S_i  =  -tr(\rho_i\log \rho_i )$, in which
$\rho_i=Tr_i\left |\Psi\right \rangle\left\langle \Psi\right |$ with $Tr_i$ standing for
the tracing over all the particles except the $i$th one. The results are shown in  Fig. \ref{es}(b). It is obvious that the optimized bases also makes the entanglement converge much faster. Compared with the ground state energy, entanglement needs more number of bases to get convergent. With the present parameters used in the calculation, the number is 10 for entanglement and 8 for the ground state energy. Anyway, both have shown us the advantaged and the necessity to use the optimized basis set.

\begin{figure}
\includegraphics[angle=-90,width=10cm]{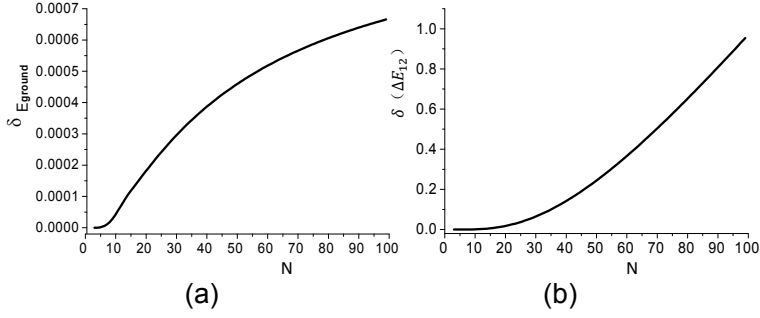}
\hfill \caption{Influence of the system size $N$ upon relative calculation error of the the ground state energy [(a)] and the energy gap between the ground state and the first excited state[(b)]. To keep the calculation amount controllable, 10 renormalized bases are used.} \label{Error}
\end{figure}

\begin{table}
\caption{ List of the first $20$ eigenvalues of the reduced density matrix calculated
by DMRG for different number of targeted states $n_{tar}$.  The number of the optimized bases is $8$.}
\begin{ruledtabular}
\renewcommand{\arraystretch}{0.6}\setlength{\LTcapwidth}{5in}
\begin{tabular}{llllll}
 $N $ &  $n_{tar}$=1   &   $n_{tar}$=2  &  $n_{tar}$=3  &  $n_{tar}$=4    &  $n_{tar}$=5    \\
\hline\hline
            1   &  0.9289161 &   0.7165831  &  0.6081884  &  0.5142144  &  0.4662047\\
            2   &  0.06598694 &   0.2520588  &  0.3114651 &   0.3423709  &  0.2923608 \\
            3   &  0.004625028  &  0.02200897 &   0.04093407  &  0.09485543  &  0.01222936  \\
            4   &  0.3072901$\times10^{-3}$ &   0.007458359  &  0.03358179  &  0.03575154 &  0.09736594 \\
            5   &  0.1380769$\times10^{-3}$  &   0.001387261  &  0.003139719  &  0.008861609  &  0.0105103  \\
            6   &  0.1762733$\times10^{-4}$  &   0.3988337$\times10^{-3}$  &  0.002124129  &  0.002502725  &  0.006969623  \\
            7   &  0.8231959$\times10^{-5}$  &  0.7059047$\times10^{-4}$ &   0.2318205$\times10^{-3}$  &  0.7169158$\times10^{-3}$  &  0.002355332  \\
            8   &  0.7004932$\times10^{-6}$  &  0.2112216$\times10^{-4}$  &  0.1969156$\times10^{-3}$ &   0.3126910$\times10^{-3}$   &  0.001035142  \\
            9   &  0.2749973$\times10^{-17}$  &   0.5958590$\times10^{-5}$  &  0.7897836$\times10^{-4}$  &  0.2704249$\times10^{-3}$   &  0.3736590$\times10^{-3}$    \\
           10   &  0.8871970$\times10^{-17}$   &   0.393011$\times10^{-5}$  &  0.3168917$\times10^{-4}$  &  0.7322320$\times10^{-4}$    &  0.3436789$\times10^{-3}$   \\
           11   &  0.4808106$\times10^{-17}$  &   0.1921591$\times10^{-5}$   &  0.1315241$\times10^{-4}$  &  0.3895618$\times10^{-4}$   &  0.8568700$\times10^{-4}$  \\
           12   &  0.1366569$\times10^{-17}$  &   0.7967883$\times10^{-6}$   &  0.9506539$\times10^{-5}$  &   0.1681713$\times10^{-4}$  &  0.5683950$\times10^{-4}$  \\
           13   &  0.1081680$\times10^{-17}$  &   0.1965801$\times10^{-6}$   &  0.2056333$\times10^{-5}$    &  0.9431559$\times10^{-5}$   &  0.2391172$\times10^{-4}$  \\
           14   &  0.5562934$\times10^{-18}$  &   0.9791207$\times10^{-7}$  &  0.1694848$\times10^{-5}$   &  0.2078514$\times10^{-5}$   &  0.1352827$\times10^{-4}$    \\
           15   &   0.2082794$\times10^{-18}$  &   0.6546016$\times10^{-7}$ &   0.4700800$\times10^{-6}$  &   0.1576653$\times10^{-5}$  &   0.2540277$\times10^{-5}$   \\
           16   &  0.1618244$\times10^{-18}$  &   0.1013991$\times10^{-7}$   &  0.3715984$\times10^{-6}$    &  0.5760311$\times10^{-6}$  &  0.2021552$\times10^{-5}$   \\
           17   &   0.1378375$\times10^{-18}$  &  0.2459260$\times10^{-16}$ &   0.6119707$\times10^{-7}$  &   0.4719433$\times10^{-6}$  &  0.1128699$\times10^{-5}$  \\
           18   &  0.1341955$\times10^{-18}$  &  0.1495277$\times10^{-16}$  &  0.1599546$\times10^{-7}$  &   0.1424888$\times10^{-6}$  &   0.6702679$\times10^{-6}$    \\
           19   &  0.7397374$\times10^{-19}$  &   0.4536211$\times10^{-17}$  &   0.1112929$\times10^{-7}$  &   0.6986116$\times10^{-7}$ &   0.4696767$\times10^{-6}$  \\
           20   &   0.4642735$\times10^{-19}$  & 0.4146682$\times10^{-17}$  &   0.7577889$\times10^{-9}$  &   0.3261627$\times10^{-7}$  &   0.2184679$\times10^{-6}$  \\
\end{tabular}
\end{ruledtabular}
\label{table1}
\end{table}

\begin{figure}
\includegraphics[angle=-90,width=13cm]{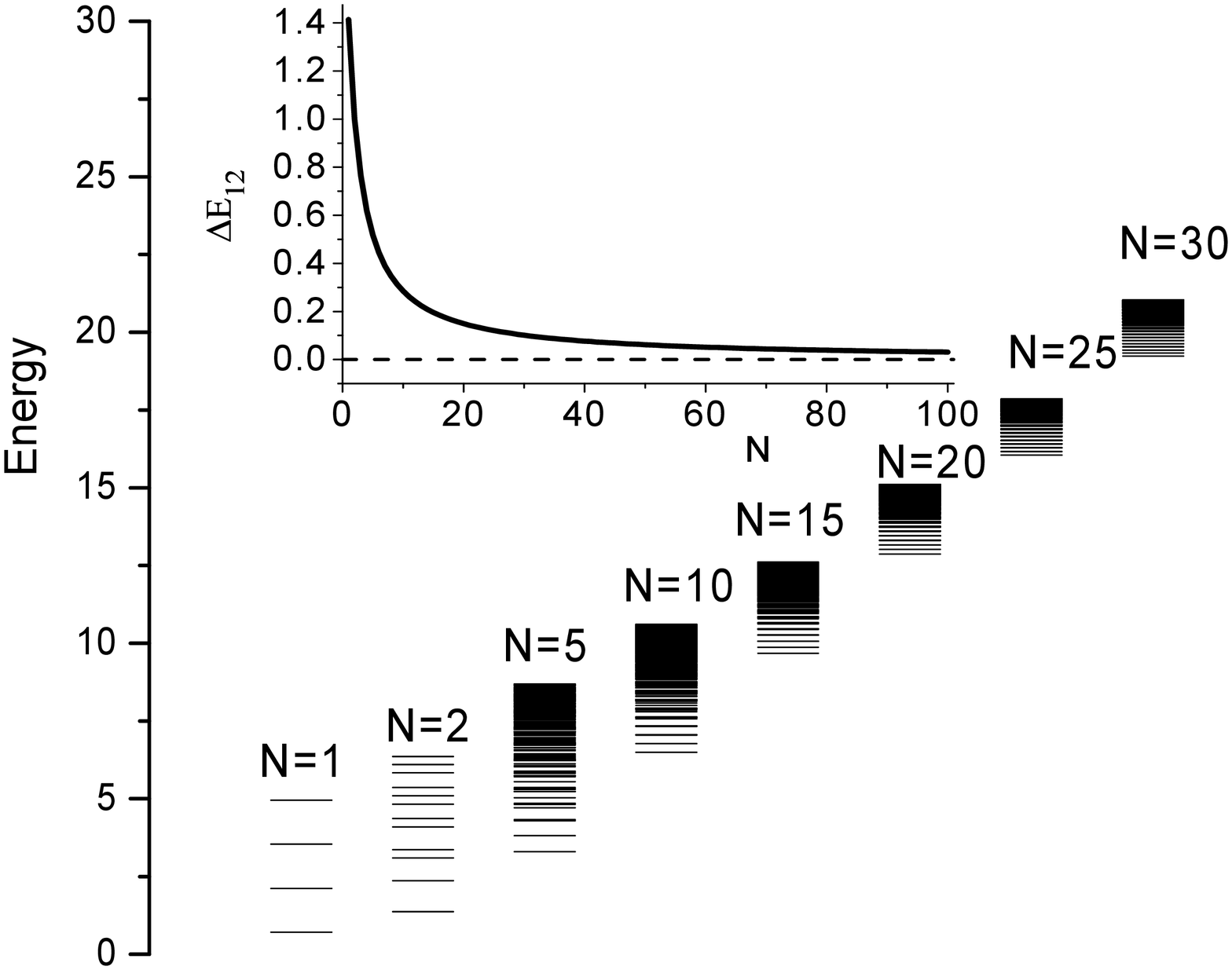}
\caption{Dependence of the energy spectrum upon the system size $N$. The inset depicts the dependence of the first energy gap
$\bigtriangleup E_{12}$ on $N$.}
\label{energyspectrum}
 \end{figure}

The next important physical quantity we will analysis is the energy gap $\Delta E_{12}$ between the ground state and the first excited state. Generally, it is much more difficult to get $\Delta E_{12}$
than $E_{ground}$, especially for larger system size. For comparison, Fig. \ref{Error} presents the error dependence of these two quantities upon the system size, from which we can observe two effects.
\par
Firstly, the calculation of the ground state energy is much more accurate than the energy gap. For example, for $\Delta E_{12}$, when $N=100$, the calculation error is already unacceptable since the relative error now is alomost $100\%$. But for $E_{ground}$, even when $N=100$, we can still get very accurate results. Here it should be noted that both quantities are calculated with the same number of optimized bases with $n=10$ in Fig. \ref{Error}. We can improve the accuracy by using bigger $n$. Normally, how many optimized bases are needed is decided by the cutoff of the eigenvalues of the reduced density matrix, i.e. $\{\lambda_i,i=1,2,\cdots,\lambda_1\geq \lambda_2 \geq \cdots\}$. Considering the condition $\sum_i \lambda_i=1$, the cutoff error can be roughly estimated as $1-\sum_{i=1}^n{\lambda_i}$. Hence, the success of the DMRG algorithm strongly depends upon descending speed of $\lambda_i (i=1,2,\cdots)$. So to check why the calculation of $\Delta E_{12}$ need more optimized bases, we must know the difference of $\{\lambda_i,i=1,2,\cdots\}$ when calculating $\Delta E_{12}$ and $E_{ground}$. As we know, the reduced density matrix is obtained from the targeted states. The number of the targeted states $n_{tar}$ is decided by the energy levels we are interested in. For example, if we are interested in calculating the ground state energy $E_{ground}$, $n_{tar}=1$. If $\Delta E_{12}$ needs to be calculated, $n_{tar}=2$ since it involves two energy levels. Usually, for different $n_{tar}$, $\{\lambda_i,i=1,2,\cdots\}$ will be different and thus the cutoff number $n$ will be different. To have some ideas about the above analysis, in Table \ref{table1}, we give the list of the first $20$ eigenvalues of the $64\times 64$ reduced density matrix for $N=50$. Assume that we only keep the eigenstates with eigenvavlues bigger than $\sim 10^{-6}$. It is interesting to note that when $n_{tar}=1$, $n=8$ is enough. But for $n_{tar}=2,3,4,5$, $n$ needs to be $12,15,16,18$, respectively. That means if we want to get the first 5 energy levels, a matrix for the system block with dimension $18^3=5832$ needs to be solve repeatedly in the DMRG algorithm, which is really a tremendous burden for the computer. In fact, in our work, to save the computer time, we have used just one free site between the blocks. If two free sites are used as in the conventional work, we will be challenged by solving a matrix with dimension $18^4=104976$, which will makes the calculation an impossible task. From these discussions, we can see that the number of the targeted states is another important source to influence the algorithm efficiency.

Secondly, Fig. \ref{Error} also demonstrates the big influence of the chain size upon the calculation errors, which increases quickly
with the chain size, especially for $\Delta E_{12}$. The reason can be attributed to the structure of the low-lying energy levels of the system. To clarify this point, we
plot the energy spectrum in Fig. \ref{Error}. One obvious trend of the spectrum is that more and more energy levels are emerging and the the energy level spacing decreases very quickly with the size. For more clear demonstration, we also draw $\Delta E_{12}$ as a function of $N$ in the inset of Fig. \ref{Error}. The consequence of the decreasing energy level spacing is that the higher energy levels will be unavoidably mixed with or influence the truncated Hilbert space. Then $\{\lambda_i,i=1,2,\cdots\}$ will decrease more and more slowly, which will leads to the increased number of optimized states. That is exactly the reason for the low accuracy of $\Delta E_{12}$ if we keep $n$ fixed while increasing the system size.

\par
 In summary, some important information upon the error sources and efficiency improvements of DMRG algorithm is provided in this letter by using the harmonic oscillator chain as an example. Firstly, the usage of optimized bases is a necessity for a bosonic system. Secondly, the number of targeted states will severely influence the accuracy of the results. Thirdly, the energy structure of the whole system also plays a key role in justifying the use of DMRG method. It is more suitable for a system with bigger energy gap. According to our experience with other models, such as quantum Frenkel-Kontorova model \cite{F-K4,wang2007}, these conclusions are not just limited to the harmonic oscillator chain, they are having more general significance to guide us in applying
 this powerful numerical algorithm.

This work is supported by the National Natural Science Foundation of China under Grant Nos. 11274117 and Shanghai Excellent academic leaders Program of China (Grant No. 12XD1402400).


\begin{thebibliography}{40}

\bibitem{dmrg01}  White S R 1992 Phys. Rev. Lett. {\bf 69}  2863
\bibitem{dmrg1}   White S R 1993  Phys. Rev. B. {\bf 48}  10345
\bibitem{dmrg2}   White S R, Feiguin A E 2004 Phys. Rev. Lett. {\bf 93}  076401
\bibitem{F-K4}    Hu B and Wang J 2006  Phys. Rev. B {\bf 73}  184305
\bibitem{dmrg0}   Caron L G and Moukouri S 1997  Phys. Rev. B. {\bf 56}  8471

\bibitem{method1}  Weisse A, Fehske H, Wellein G and Bishop A R 2000 Phys. Rev. B. {\bf62}  747
\bibitem{method2}  Weiss A, Wellein G  and Fehske H 2002 High Performance Computing in Science and Engineering (springer, Berlin) vol 02  p131


\bibitem{method3}  Zhang C, Jeckelmann E and White S R 1998  Phys. Rev. Lett. {\bf 80}  2661
\bibitem{method4}  Friedman B 2000 Phys. Rev. B. {\bf 61} 6701

\bibitem{entangle1} Zanardi P 2002  Phys. Rev. A. {\bf 65} 042101
\bibitem{entangle2} Wang X 2001 Phys.Rev. A. {\bf 64}  012313
\bibitem{entangle3} Gu S, Deng S, Li Y and Lin H 2004  Phys. Rev. Lett. {\bf 93} 086402

\bibitem{wang2004} Wang J and Kais S 2004 Phys. Rev.  A {\bf 70}  022301
\bibitem{wang2003} Wang J and Kais S 2003   Int. J. Quant. Infor. {\bf 1} 375
\bibitem{Kais}     Kais S 2007 Adv. Chem. Phys. {\bf 134}, 493
\bibitem{wang2007} Wang J, Hu B and Wang X 2007  Prog. Theo. Phys., Suppl. {\bf 166}, 95





\end{thebibliography}
\end{document}